# Coherent States of Multilayer Lithiated Graphene utilizing an Electric Vector Potential

Donald C. Boone
Nanoscience Research Institute, Arlington, VA 22314 USA; db2585@caa.columbia.edu

**Abstract: The computational research that will be presented compares the coherent states of multiple layer graphene versus the coherent states of lithium ions diffused within this multilayer graphene. Unlike the prevailing research on graphene coherent states that uses an external magnetic vector potential, this research will utilize an electric vector potential that is inherent from the electron current density that transverses the π-bonds on the graphene surface. The results from this study will display that the electric vector potential produces energies that can lead to coherent states for graphene and lithiated graphene without the necessity of a strong applied magnetic vector potential. Two excited states will be analyzed for each material. The first excited state (n=1) the graphene coherent state energies lags behind the energies of the lithiated graphene coherent state where as the second excited state (n=2) the coherent state energies of graphene exceeds that of the lithiated graphene coherent state.**

## Introduction:

Graphene is an allotrope of carbon in which the carbon atoms are arranged in a two dimensional (2D) hexagonal lattice that is one atom thick. When an electromagnetic potential is applied to the 2D graphene surface it becomes an above average conductor of electrons that are best described as massless Dirac fermions. When these electrons are in the vicinity of the hexagonal lattice points called Dirac points the electrons travel at a constant speed called the Fermi velocity $v_F \sim 10^6$m/s which is analogous to the massless photon speed of light c. Therefore the effective mass of the electrons are theoretically zero on the graphene surface near the Dirac points. Because of this analogy between photons and massless electrons on the graphene surface, the relativistic Dirac-Weyl equation (1) will be used to describe the electric and magnetic fields that are produced on the graphene sheet.

$$v_F \vec{\sigma}_i \cdot \left( \vec{p} + \frac{e\,\vec{A}}{c} \right) \Psi = \lambda \Psi \qquad (1)$$

where the $\vec{\sigma}_i$ are the Pauli matrices, $\vec{p}$ is the momentum operator, **e** is the electric charge constant, **c** is the speed of light and $\vec{A}$ is the component of the electromagnetic vector potential in the orthogonal direction to the graphene surface. The Dirac-Weyl equation will be solved as an eigenvalue equation with the solution of the eigenvalues λ representing the energy states composed of the π-bonds of the $sp^2$ carbon atom orbitals. The π-bonds are the reason for the high conductivity of graphene due to the movement of electrons through these discrete π-bonds energy states. The solutions to the eigenstates Ψ are the wavefunctions representing the electromagnetic fields in the ground and excited state. In order to solve the Dirac-Weyl equation the electromagnetic field is decomposed into the electric and magnetic fields and each are solved separately as an energy eigenvalue equation. [1]

In this research the two geometric models that will be studied are the three dimensional multilayer lithiated graphene (which will be defined as lithiated graphite) and the single layer sheet of two dimensional graphene. The 2D graphene will be analysis with an effective electron mass of zero ($m_e$=0) for reasons stated in the beginning of this introduction however, the multilayer lithiated graphene or lithiated graphite has a non-zero effective mass and therefore this model will be analysis with an effective electron mass $m_e$>0.

## Electric and Magnetic Vector Potentials:

The electromagnetic vector potential $\vec{A}$ will be decoupled into the electric vector potential $\vec{A}_{\vec{E}_n}$ and magnetic vector potential $\vec{A}_{\vec{B}_n}$. The following equations in this section are derived from the work of Bautista et al. [2]. The set of equations used to solve for the electric field energy eigenvalues $\lambda_{\vec{E}_n}$ and eigenstates $\Psi_{\vec{E}_n}$ are

$$v_F \vec{\sigma}_i . \left( \vec{p} + \frac{e\vec{A}_{\vec{E}_n}}{c} \right) \Psi_{\vec{E}_n}(\vec{r}) = \lambda_{\vec{E}_n}(\vec{r}) \Psi_{\vec{E}_n}(\vec{r}) \quad (2)$$

$$\vec{A}_{\vec{E}_n}(\vec{r}) = \frac{e}{4\pi\varepsilon_o a^3} \int_V \frac{\left|\chi_{\vec{E}_n}(\vec{r})\right|^2}{|\vec{r} - \vec{r}_o|} d^3\vec{r}_o \quad (3)$$

For the ground state (n=0)

$$\chi_{\vec{E}_n}(\vec{r}) = 2\vec{r}^{\frac{3}{2}} \left(\frac{\alpha_{\vec{E}}}{2\pi}\right)^{\frac{3}{2}} exp\left[ -\frac{\alpha_{\vec{E}}{}^2}{4} \left( \vec{r} + \frac{2\vec{k}_z}{\alpha_{\vec{E}}{}^2} \right)^2 \right] \quad (4)$$

and for the excited state n=1,2

$$\chi_{\vec{E}_n}(\vec{r}) = \vec{r}^{\frac{3}{2}} \left(\frac{n+2}{2n}\right) \left(\frac{\alpha_{\vec{E}}}{2\pi}\right)^{\frac{3}{2}} exp\left[ -\frac{\alpha_{\vec{E}}{}^2}{4} \left( \vec{r} + \frac{2\vec{k}_z}{\alpha_{\vec{E}}{}^2} \right)^2 \right] \quad (5)$$

$$\alpha_{\vec{E}} = \left(\frac{e\vec{E}}{c\hbar}\right)^{\frac{1}{2}} \quad (6)$$

where $\chi_{\vec{E}_n}$ is the graphene $\pi$-bond wavefunction for the electric field. The graphene electric field $\vec{E}$ is

$$\vec{E} = \vec{F}_{\vec{E}} \pm \frac{v_F \mu_o \left[ 2E(\vec{k})\varepsilon_o v_F^2 + 2E(\vec{k})\mu_o{}^{-1} - \varepsilon_o \mu_o{}^{-1} \vec{F}_{\vec{E}}{}^2 \right]^{1/2}}{\varepsilon_o \mu_o v_F^2 + 1} \quad (7)$$

where $\vec{F}_{\vec{E}}$ is the applied electric field that initiates the flow of electrons that travel the 2D graphene lattice.

As much the same way equations 3 thru 7 are for the electric vector potential $\vec{A}_{\vec{E}_n}$, equations 9 thru 13 are the analogous version for the magnetic vector potential $\vec{A}_{\vec{B}_n}$. The set of equations used to solve for the magnetic field energy eigenvalues $\lambda_{\vec{B}_n}$ and eigenstates $\Psi_{\vec{B}_n}$ are

$$v_F \vec{\sigma}_i . \left( \vec{p} + \frac{e\vec{A}_{\vec{B}_n}}{c} \right) \Psi_{\vec{B}_n}(\vec{r}) = \lambda_{\vec{B}_n}(\vec{r}) \Psi_{\vec{B}_n}(\vec{r}) \quad (8)$$

$$\vec{A}_{\vec{B}_n}(\vec{r}) = \frac{e\mu_o v_F^2}{4\pi a^3} \int_V \frac{\left|\chi_{\vec{B}_n}(\vec{r})\right|^2}{|\vec{r} - \vec{r}_o|} d^3\vec{r}_o \quad (9)$$

For n=0 ground state

$$\chi_{\vec{B}_n}(\vec{r}) = 2\vec{r}^{\frac{3}{2}} \left(\frac{\alpha_{\vec{B}}}{2\pi}\right)^{\frac{3}{2}} exp\left[ -\frac{\alpha_{\vec{B}}{}^2}{4} \left( \vec{r} + \frac{2\vec{k}_z}{\alpha_{\vec{B}}{}^2} \right)^2 \right] \quad (10)$$

and for n=1,2 excited state

$$\chi_{\vec{B}_n}(\vec{r}) = \vec{r}^{\frac{3}{2}}\left(\frac{n+2}{2n}\right)\left(\frac{\alpha_{\vec{B}}}{2\pi}\right)^{\frac{3}{2}} exp\left[-\frac{\alpha_{\vec{B}}{}^{2}}{4}\left(\vec{r}+\frac{2\vec{k}_z}{\alpha_{\vec{B}}{}^{2}}\right)^{2}\right] \tag{11}$$

$$\alpha_{\vec{B}} = \left(\frac{e\vec{B}}{\bar{h}}\right)^{\frac{1}{2}} \tag{12}$$

where $\chi_{\vec{B}_n}$ is the graphene π-bond wavefunction for the magnetic field. The graphene magnetic field $\vec{B}$ is

$$\vec{B} = \pm\frac{\mu_o\left[\left(2E(\vec{k})\varepsilon_o v_F^2 + 2E(\vec{k})\mu_o{}^{-1} - \varepsilon_o\mu_o{}^{-1}\vec{F}_{\vec{E}}{}^{2}\right)^{1/2} \pm \vec{F}_{\vec{E}}\varepsilon_o v_F\right]}{\varepsilon_o\mu_o v_F^2 + 1} \tag{13}$$

The electric and magnetic fields were derived from the Lorentz force and electromagnetic field energy density equations.

For graphene the energy dispersion relation $E(\vec{k})$ is defined as

$$E(\vec{k}) = \pm\gamma_\pi\left(1 + 4\cos\frac{\sqrt{3}}{2}ak_x\cos\frac{a}{2}k_y + 4\cos^2\frac{a}{2}k_y\right)^{1/2} \tag{14}$$

while the energy dispersion relation $E(\vec{k})$ for lithiated graphene is defined as

$$E(\vec{k}) = \frac{\bar{h}^2}{m_e}\left(k_x^2 + k_y^2 + k_z^2\right) \tag{15}$$

where $a$ is the graphene lattice constant, $m_e$ is the effective electron mass of graphite [3], Planck constant is $\bar{h}$, the $\gamma_\pi$ energy parameter is defined from experimental data [4], and the wave vectors $k_x$, $k_y$, $k_z$ for the 2D graphene surface of x and y directions and three dimensional lithiated graphite volume for the x,y,z directions.

The calculations of these two sets of equations has discovered that the electric vector potential $\vec{A}_{\vec{E}_n}$ is of a magnitude far greater than that of magnetic vector potential $\vec{A}_{\vec{B}_n}$ due to the fact that graphene is essentially a non-magnetic material. However, with an adequate applied electric field $\vec{F}_{\vec{E}}$ , the electric vector potential can be used as an effective energy source orthogonal to the graphene surface that will produce a coherent state.[5]

The Dirac-Weyl equation $H\Psi = \lambda\Psi$ introduced in equation (1) in matrix form is

$$\begin{bmatrix} H_{\vec{E}_n} & H_{\vec{B}_n} \\ H_{\vec{E}_0} & H_{\vec{B}_0} \end{bmatrix}\begin{Bmatrix} \Psi_{\vec{E}_n} \\ \Psi_{\vec{B}_n} \end{Bmatrix} = \lambda_{0,n}\begin{Bmatrix} \Psi_{\vec{E}_n} \\ \Psi_{\vec{B}_n} \end{Bmatrix} \tag{16}$$

with individual elements of the Hamiltonian $H$ defined in equations (17a-d) as

$$H_{\vec{E}_0} = v_F\vec{\sigma}_i.\left(\vec{p}_{\vec{E}_0} + \frac{e\vec{A}_{\vec{E}_0}}{c}\right) \tag{17a}$$

$$H_{\vec{B}_0} = v_F\vec{\sigma}_i.\left(\vec{p}_{\vec{B}_0} + \frac{e\vec{A}_{\vec{B}_0}}{c}\right) \tag{17b}$$

$$H_{\vec{E}_n} = v_F\vec{\sigma}_i.\left(\vec{p}_{\vec{E}_n} + \frac{e\vec{A}_{\vec{E}_n}}{c}\right) \tag{17c}$$

$$H_{\vec{B}_n} = v_F\vec{\sigma}_i.\left(\vec{p}_{\vec{B}_n} + \frac{e\vec{A}_{\vec{B}_n}}{c}\right) \tag{17d}$$

where n=1,2 are the excited states within the Hamiltonian H. The solution to equation (16) gives the eigenstates $\Psi_{\vec{E}_0}$ and $\Psi_{\vec{B}_0}$ in which the ground state wave function $\psi_g$ is calculated as the product of $\Psi_{\vec{E}_0}$ and $\Psi_{\vec{B}_0}$

$$\psi_g = \langle\Psi_{\vec{E}_0}\,|\Psi_{\vec{B}_0}\rangle = -\langle\Psi_{\vec{B}_0}\,|\Psi_{\vec{E}_0}\rangle \tag{18}$$

**Coherent States:**

After defining the energy dispersion relation $E(\vec{k})$, the Hamiltonian $H$ and ground state wave function $\psi_g$, a series of equations are used to calculate the optical amplification of the electromagnetic field in order to determine the graphene and lithiated graphene coherent states.[6]

$$\omega_p = \frac{1}{\hbar}\langle\psi_g|H|\psi_g\rangle \tag{19}$$

$$\psi_n = \frac{\langle\psi_m|H|\psi_g\rangle}{E(\vec{k}) - E_0}\psi_m \tag{20}$$

$$A_{21} = \frac{4e^2\omega_p^3}{3\varepsilon_o\hbar c^3}\left|\langle\psi_g|r|\psi_n\rangle\right|^2 \tag{21}$$

$$\varepsilon_r = -\frac{\varepsilon_o\vec{E}^2}{\left(\frac{\vec{B}^2}{\mu_o} - \frac{2m_e v_F^2}{a^3}\right)} \tag{22}$$

$$\omega_\gamma = \frac{a^3}{2\hbar}\left[\varepsilon_r\vec{E}^2 + \frac{1}{\mu_o}\vec{B}^2\right] \tag{23}$$

$$g(\omega) = \frac{\omega_p - \omega_\gamma}{2\pi}\left[{\omega_\gamma}^2 + \left(\frac{\omega_p - \omega_\gamma}{2}\right)^2\right]^{-1} \tag{24}$$

$$\sigma_{21} = A_{21}\frac{l^2}{8\pi\varepsilon_r}g(\omega) \tag{25}$$

$$\gamma = (\mathrm{r},\mathrm{t}) = \sigma_{21}(N_2 - N_1) \tag{26}$$

$$\bar{n} = \frac{(\vec{E}\times\vec{B})a^3}{\mu_o\hbar\omega_\gamma c}\mathrm{e}^{\gamma\mathrm{r}} \tag{27}$$

$$\mathrm{P} = \left[\frac{\bar{n}^{\,n}exp(-\bar{n})}{n!}\right]^{\frac{1}{2}} \tag{28}$$

In equation 19 the perturbed angular frequency $\omega_p$ of the electromagnetic field is defined as the expectation value of the Hamiltonian while existing in the groundstate $\psi_g$. The excited state wavefunction $\psi_n$ is define in equation 20 where $\psi_m$is a wavefunction that is orthogonal to $\psi_n$ and $E_0$ is the groundstate energy of the electromagnetic energy system. Equation 21 is the Einstein Coefficient $A_{21}$where $\varepsilon_o$ is the electric permittivity. The relative permittivity $\varepsilon_r$ is define in equation 22 with $\mu_o$ as the magnetic permeability. Equation 23 is the electromagnetic angular frequency $\omega_\gamma$while equations 24 thru 28 are the set of equations that simulates the stimulated emission process that leads to coherent states of graphene and lithiated multilayer graphene, $P_G$ and $P_{LG}$ respectively in equation 28.

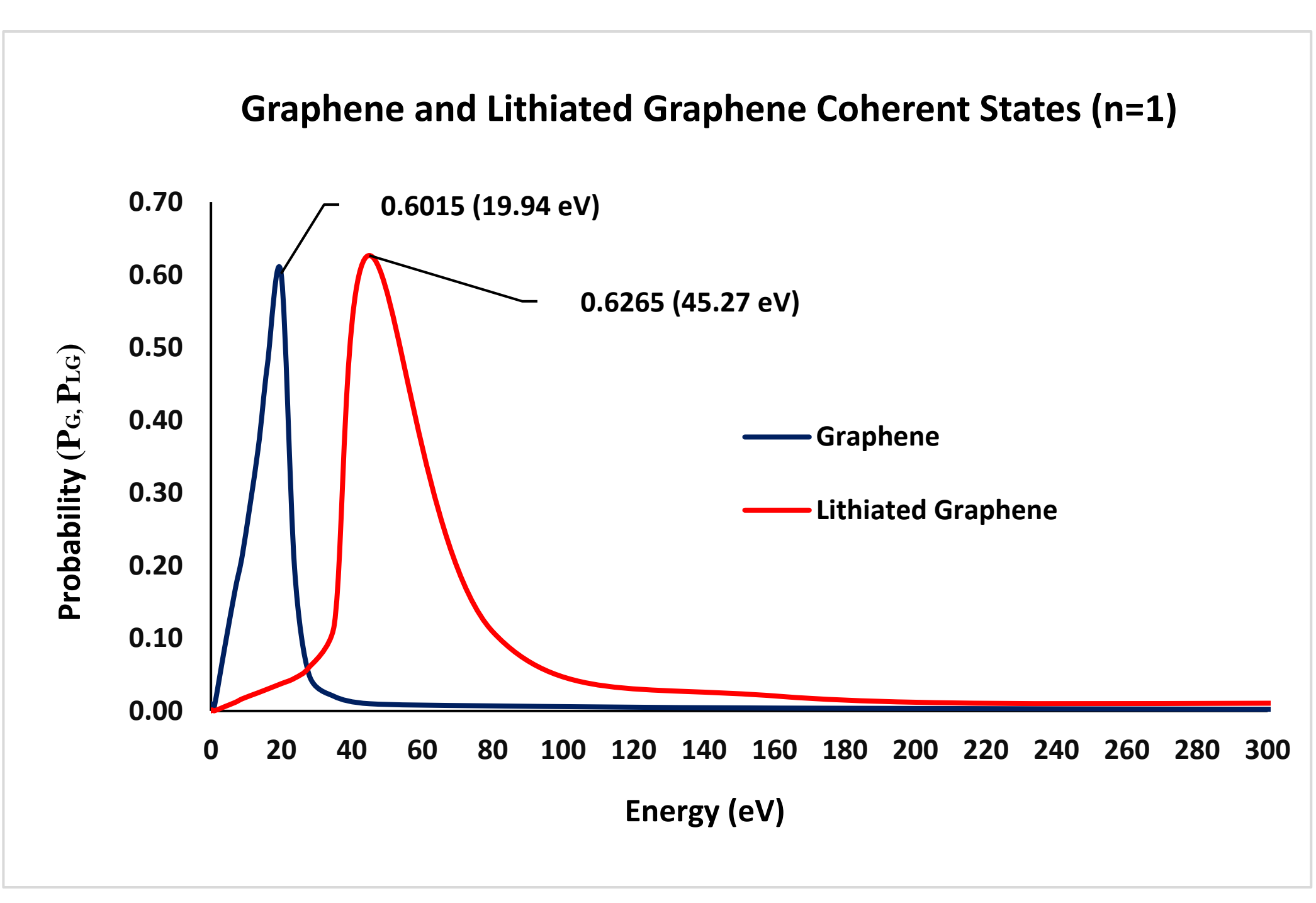

**Figure 1.** The coherent state of graphene (blue) lags behind the coherent state of lithiated graphene (red) when the excited state is n=1.

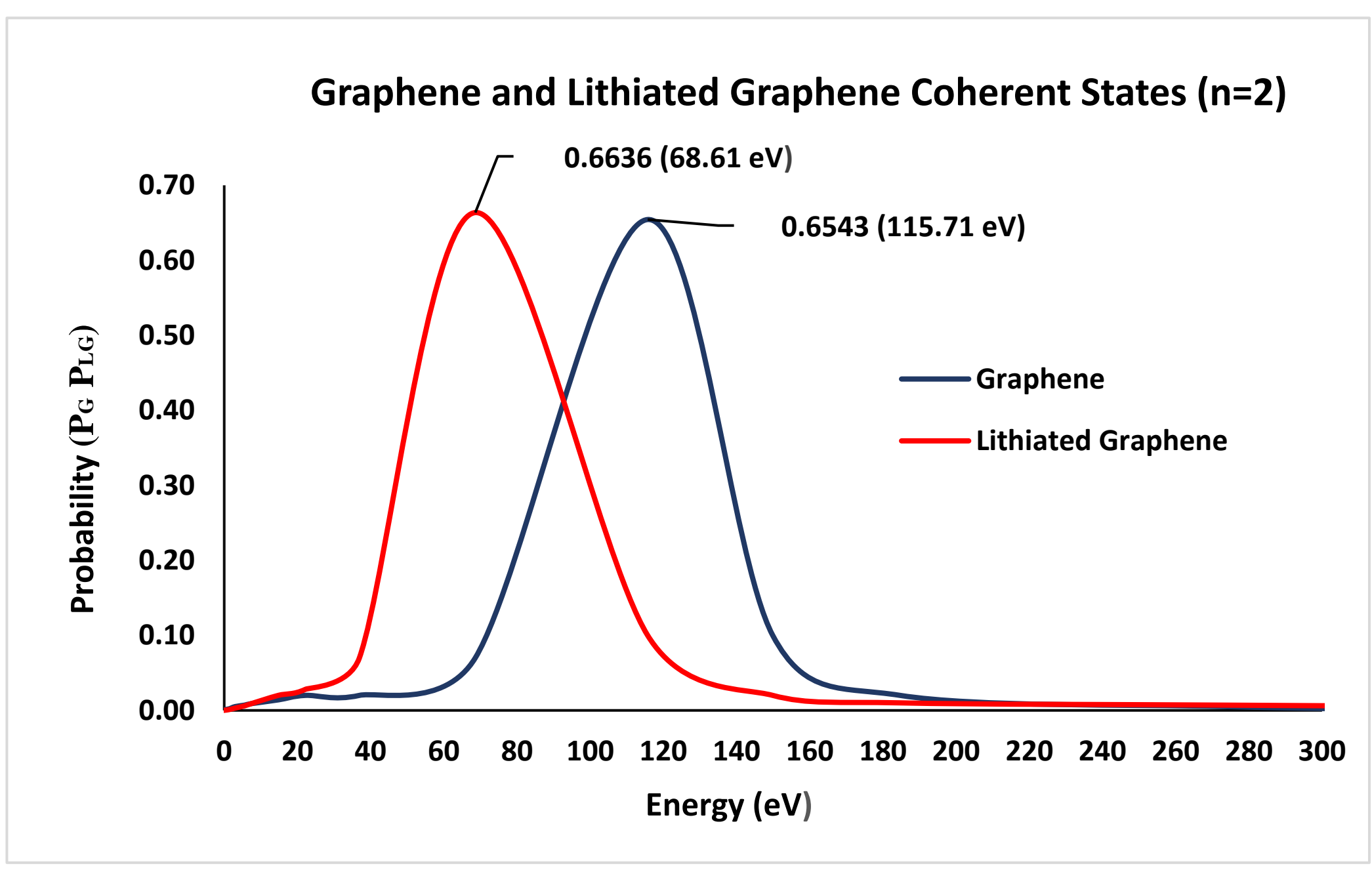

**Figure 2.** The coherent state of lithiated graphene (red) lags behind the coherent state of graphene (blue) when the excited state is n=2.

## Results:

As was displayed in figure 1, the first excited state (n=1) the graphene coherent state energies lags behind the energies of the lithiated graphene coherent state. The opposite occurs when the coherent state energies of graphene exceeds that of lithiated graphene coherent state for the second excite state (n=2) in figure 2.

## Summary:

The results from this study has displayed that the electric vector potential produces energies that can lead to coherent states for graphene and lithiated graphene without the use of a strong applied magnetic vector potential.